\documentclass[letterpaper, 10 pt, conference]{ieeeconf}  % Comment this line out if you need a4paper

\IEEEoverridecommandlockouts                              % This command is only needed if 
\usepackage{graphics} % for pdf, bitmapped graphics files
\usepackage{epsfig} % for postscript graphics files
\usepackage{amsmath} % assumes amsmath package installed
\usepackage{amssymb}  % assumes amsmath package installed

\usepackage{hyperref}
\usepackage{xcolor}
\usepackage{graphicx}
\usepackage{subcaption}
\usepackage{ntheorem}

\newtheorem{definition}{Definition}
\newtheorem{lemma}{Lemma}

\usepackage{cancel}

\usepackage{etoolbox}

\makeatletter
\let\nobreakitem\item
\let\@nobreakitem\@item
\patchcmd{\nobreakitem}{\@item}{\@nobreakitem}{}{}
\patchcmd{\nobreakitem}{\@item}{\@nobreakitem}{}{}
\patchcmd{\@nobreakitem}{\@itempenalty}{\@M}{}{}
\patchcmd{\@xthm}{\ignorespaces}{\nobreak\ignorespaces}{}{}
\patchcmd{\@ythm}{\ignorespaces}{\nobreak\ignorespaces}{}{}

\renewtheoremstyle{break}%
  {\item{\theorem@headerfont
          ##1\
          ##2\theorem@separator}\hskip\labelsep\relax\nobreakitem}%
  {\item{\theorem@headerfont
          ##1\ ##2\ (##3)\theorem@separator}}%\hskip\labelsep\relax\nobreakitem}
\makeatother

\theoremheaderfont{\kern-0cm\normalfont\bfseries} 
\theoremstyle{break}
\newtheorem{theorem}{Theorem}

\usepackage{cite}
\makeatletter 
\pretocmd\@bibitem{\color{black}\csname keycolor#1\endcsname}{}{\fail}
\newcommand\citecolor[1]{\@namedef{keycolor#1}{\color{blue}}}
\makeatother
\title{\LARGE \bf
Continuification control of large-scale multiagent systems under limited sensing and structural perturbations
}

\author{Gian Carlo Maffettone$^{1}$, Maurizio Porfiri$^{2, \dagger, *}$, Mario di Bernardo$^{1, 3, \dagger, *}$% <-this % stops a space
\thanks{This work has been partially supported by the National Science Foundation Grant No. CMMI-1932187 and by the Research Project SHARESPACE funded by the European Union (EU HORIZON-CL4-2022-HUMAN-01-14. SHARESPACE. GA 101092889 - http://sharespace.eu). Views and opinions expressed are however those of the author(s) only and do not necessarily reflect those of the European Union. The European Union cannot be held responsible for them.}%
\thanks{$^{1}$Scuola Superiore Meridionale, Naples, Italy}%
\thanks{$^{2}$ Center for Urban Science and Progress, Department of Biomedical Engineering, Department of Mechanical and Aerospace Engineering, Tandon School of Engineering, New York University, USA}%
\thanks{$^{3}$ Department of Electric Engineering and Information Technology, University of Naples Federico II, Naples, Italy}%
\thanks{$\dagger$ These authors contributed equally}
\thanks{$^{*}$For correspondence: {\tt\small mario.dibernardo@unina.it, mporfiri@nyu.edu}}%
}
\usepackage[T1]{fontenc}

\begin{document}

\maketitle
\thispagestyle{empty}
\pagestyle{empty}

%%%%%%%%%%%%%%%%%%%%%%%%%%%%%%%%%%%%%%%%%%%%%%%%%%%%%%%%%%%%%%%%%%%%%%%%%%%%%%%%
\begin{abstract}
We investigate the stability and robustness properties of a continuification-based strategy for the control of large-scale multiagent systems. Within continuation-based strategy, one transforms the microscopic, agent-level description of the system dynamics into a macroscopic continuum-level, for which a control action  can be synthesized to steer the macroscopic dynamics towards a desired distribution. Such an action is ultimately discretized to obtain a set of deployable control inputs for the agents to achieve the goal. The mathematical proof of convergence toward the desired distribution typically relies on the assumptions that no disturbance is present and that each agent possesses global knowledge of all the others' positions. Here, we analytically and numerically address the possibility of relaxing these assumptions for the case of a one-dimensional system of agents moving in a ring. We offer compelling evidence in favor of the use of a continuification-based strategy when agents only possess a finite sensing capability and spatio-temporal perturbations affect the macroscopic dynamics of the ensemble. We also discuss some preliminary results about the role of an integral action in the macroscopic control solution.
\end{abstract}

%%%%%%%%%%%%%%%%%%%%%%%%%%%%%%%%%%%%%%%%%%%%%%%%%%%%%%%%%%%%%
%%%%%%%%%%%%%%%%%%%%%%%%%%%%%%%%%%%%%%%%%%%%%%%%%%%%%%%%%%%%%

\section{Introduction}
% What to say:
% \begin{itemize}
%     \item Recall continuification approach (ref to figure from previous paper)
%     \item Highlight criticism to continuification control (sensing capabilities is probably the most relevant)
%     \item Say that we consider the framework of \cite{maffettone2022continuification}, but considering a control action that is able to guarantee GAS in the nominal case (non la versione lineare nell'errore che presentavamo nell'altro paper).
%     \item organization of the work
% \end{itemize}
%\IEEEPARstart{C}
Continuification (or continuation) control was first proposed in \cite{nikitin2021continuation} as a viable approach to control the collective behavior of large-scale multiagent systems. %{\color{blue} [Ho messo come refs Dome e due papers (di cui uno di Davide S. che mi sembravano adeguati, ma non sono sicurissimo)]}
%and human networks \cite{Shahal2020a}.
The key idea of continuification consists of three fundamental steps: (i) finding a macroscopic description (typically a partial differential equation, PDE) for the collective dynamics of the multiagent system of interest; (ii) designing a macroscopic control action to attain the desired collective response; (iii) discretize the macroscopic control action to obtain feasible control inputs for the agents at the microscopic level.

This methodology tackles problems in which the control goal is formulated at the macroscopic collective dynamics level, but control actions are ultimately exerted only at the microscopic agent scale \cite{diBernardo2020}. Applications of the approach are related, but not limited to, multi-robot systems \cite{freudenthaler2020pde, biswal2021decentralized, qi2014multi, sinigaglia2022density}%\cite{wang2020shape, Gardi2022, giusti2022distributed, petersen2019review, zhang2022aerial}
, cell populations \cite{guarino2020balancing, agrawal2019vitro, rubio2022open}, neuroscience \cite{menara2022functional, noori2020activity}, and human networks \cite{shahal2020synchronization, calabrese2021spontaneous}.%\cite{guarino2020balancing, agrawal2019vitro, rubio2022open}
%\cite{Hauert2009a, Rubenstein2014, Gardi2022, giusti2022distributed},
%{\color{blue}[Ho messo anche citazione del nostro preprint]}

Such an approach was used in \cite{maffettone2022continuification} to control the distribution of a multiagent system swarming in a ring, leading to an effective control scheme for the multiagent system to achieve a desired distribution. Crucially, to prove convergence of the macroscopic collective dynamics towards the desired distribution, two key assumptions were made. First, that agents possess unlimited sensing capabilities so as to know the positions of all other agents in the swarm. Second, that no disturbance or perturbation is affecting the agents dynamics.

The aim of this paper is to remove these unrealistic assumptions and study the performance, stability, and robustness of the continuification approach in the presence of limited sensing capabilities of the agents, spatio-temporal disturbances,  or perturbations of their interaction kernel. In particular, we prove that local asymptotic or bounded convergence can still be achieved under these circumstances. As we undertake this task, we offer insight into the role of 
the control parameters on the size and shape of the region of asymptotic stability (or basin of attraction) of the desired distribution.

After providing some useful notation and mathematical preliminaries in Section \ref{sec:math_prel}, we briefly recall the approach of \cite{maffettone2022continuification} in Section \ref{sec:cont_control}. Then, we assess the robustness properties of the continuification control approach in Sections \ref{sec:lim_sens} and \ref{sec:struc_pert}. Finally, in Section \ref{sec:integral}, we show some preliminary results on the use of an additional integral action at the macroscopic level to improve the robustness of the microscopic dynamics, in the presence of perturbations or limited sensing.
%{\color{blue} We also show some preliminary results about modifying the proposed control action with an extra integral term. Although this is the object of ongoing work, we briefly highlight how this can easily increase robustness properties.}
Theoretical results are complemented by numerical simulations.

\section{Mathematical preliminaries}\label{sec:math_prel}
Here, we offer some useful notation and mathematical concepts that will be used throughout the paper.

\begin{definition}[Unit circle]\label{def:unit_circ}
    We define $\mathcal{S}:=[-\pi, \pi]$ as the unit circle. 
\end{definition}
\begin{definition}[$L^p$-norm on $\mathcal{S}$ \cite{Axler2020}]\label{def:Lp_norm}
    Given a scalar function of $\mathcal{S}$ and time $h:\mathcal{S}\times\mathbb{R}_{\geq0}\rightarrow\mathbb{R}$, we define its $L^p$-norm on $\mathcal{S}$ as
\begin{align}
    \Vert h(\cdot, t)\Vert_p := \left( \int_{\mathcal{S}} \vert h(x, t)\vert^p \,\mathrm{d}x\right)^{1/p}.
\end{align}
For $p=\infty$,
\begin{align}
    \Vert h(\cdot, t)\Vert_\infty :=\mathrm{ess} \,\mathrm{sup}_\mathcal{S} \vert h(x, t)\vert.
\end{align}
For the sake of brevity, we denote these norms as $\Vert h\Vert_p$, without explicitly indicating their time dependencies. 
\end{definition}
\begin{lemma} [Holder's inequality \cite{Axler2020}]\label{th:holder}
    Given $n$ $L^p$ functions, $f_i\in L^p$, with $i=1, 2, \dots n$,
    \begin{align}
        \left\Vert \prod_{i=1}^n f_i\right\Vert_1 \leq \prod_{i=1}^n \Vert f_i\Vert_{p_i}, \;\;\mathrm{if } \;\; \sum_{i=1}^n \frac{1}{p_i} = 1.
    \end{align}
    For instance, if $n=2$, we have $\Vert f_1f_2\Vert_1\leq \Vert f_1\Vert_2 \Vert f_2\Vert_2$, as well as $\Vert f_1f_2\Vert_1\leq \Vert f_1\Vert_1 \Vert f_2\Vert_\infty$.
\end{lemma}
% \begin{theorem}[Minkowsky inequality]\label{th:Minkowsky}
%     Given two $L^p$ functions, $f$ and $g$, the following inequality holds:
%     \begin{align}
%         \Vert f + g \Vert_p \leq \Vert f\Vert_p + \Vert g\Vert_p,
%     \end{align}
%     for $1\leq p\leq \infty$. This proves $L^p$ spaces are normed vector spaces, as it generalize the triangular inequality to $L^p$ functions.
% \end{theorem}
% \begin{lemma}\label{lemma:lem1}
% Given Theorem \ref{th:Minkowsky}, also the following inequality holds
% \begin{align}
%     \Vert f - g \Vert_p \leq \Vert f\Vert_p + \Vert g\Vert_p,
% \end{align}
% for $1\leq p\leq \infty$.
% \end{lemma}
We denote with \textquotedblleft{} $*$ \textquotedblright{} the convolution operator. When referring to periodic domains and functions, the operator needs to be interpreted as a circular convolution \cite{jeruchim2006simulation}. 
\begin{lemma} [Young's convolution inequality \cite{Axler2020}] \label{th:young_inequality}
    Given two functions, $f\in L^p$ and $g \in L^q$, 
    \begin{align}
        \Vert f * g\Vert_r \leq \Vert f\Vert_p \,\Vert g\Vert_q,\;\;\mathrm{if }\;\;\frac{1}{p} + \frac{1}{q} = \frac{1}{r} + 1,
    \end{align}
    where $1\leq p,q,r \leq\infty$. For instance, $\Vert f*g \Vert_\infty \leq \Vert f \Vert_2\Vert g \Vert_2$. 
\end{lemma}
We denote with the subscripts $t$ and $x$ time and space partial differentiation, respectively. 
%\begin{definition}\label{lem:young_deriv}
It can be shown \cite{jeruchim2006simulation} that the derivative of the convolution of two functions $(f*g)(x)$, can be computed as
\begin{align}
    (f*g)_x(x)  = (f_x*g)(x) = (f*g_x)(x).
\end{align}
\begin{lemma}[Comparison lemma \cite{khalil2002nonlinear}]\label{lemma:comparison_lemma}
    Given  a scalar ordinary differential equation (ODE) $v_t = f(t, v)$, with $v(t_0) =
    v_0$, where $f$ is continuous in $t$ and locally Lipschitz in $v$, if a scalar function $u$ fulfills the differential inequality
    \begin{align}
        u_t \leq f(t, u(t)), \;\;u(t_0)\leq v_0,
    \end{align}
    then
    \begin{align}
        u(t)\leq v(t),\;\;\forall\,t\geq t_0.
    \end{align}
\end{lemma}
% \begin{definition}\label{def:lie_derivative}
% Given a vector field $h: \mathbb{R}^n \rightarrow \mathbb{R}^n$ and a scalar function $\sigma:\mathbb{R} \rightarrow \mathbb{R}^n$, we define the Lie derivative of $\sigma$ with respect to $h$ 
% \begin{align}
%     \mathcal{L}_h(\sigma) = \nabla \sigma \cdot h
% \end{align}
% \end{definition}
% \begin{theorem}
%     The dynamical system
%     \begin{align}
%         \begin{cases}
%             v_t = - a_1 v - a_2 w + a_3\sqrt{v},\\
%             w_t = v,
%         \end{cases}
%     \end{align}
%     with $a_1, a_3 >0$, $a_2\geq0$ and $v,w\geq 0$, exhibit the globally asymptotically stable equilibrium set $\Omega := \{v, w : \, \sigma(v,w) = v = 0\}$.
% \end{theorem}
% \begin{proof}
    
% \end{proof}

\section{Continuification control}\label{sec:cont_control} 
As in \cite{maffettone2022continuification},
%\subsection{Microscopic dynamics}
we consider a group of $N$ identical mobile agents moving in $\mathcal{S}$.
The dynamics of the $i$-th agent can be expressed as
\begin{equation}
    \dot{x}_i = \sum_{j = 1}^N f\left(\left\{x_i,x_j\right\}_{\pi}\right) + u_i
    \label{eq:themodel},
\end{equation}
where $x_i$ is the angular position of agent $i$ on $\mathcal{S}$, $\{x_i,x_j\}_{\pi}$ is the angular distance between agents $i$ and $j$ wrapped on $\mathcal{S}$, $u_i$ is the velocity control input, and $f:\mathcal{S}\to\mathbb{R}$ is a periodic velocity interaction kernel modeling pairwise interactions between the agents \cite{maffettone2022continuification}. 
%The function $f$ takes the relative angular distance between two agents and returns a velocity. 
%As in \cite{maffettone2022continuification}, we assume that $f$ is a vanishing odd function,  discontinuous at the origin, where it takes zero value, with a bounded derivative everywhere.
%As discussed in \cite{maffettone2022continuification}, the kernel can take different functional forms to model various types of interactions (purely repulsive, purely attractive, attractive at long range and repulsive at short range, repulsive at long range and attractive at short range).
%Hence, when $u_i= 0$, we can observe different emerging behaviors depending on the kind of interactions that are modeled by $f$. Specifically, \textit{spreading}, \textit{collapsing}, \textit{clustering} and, \textit{stable aggregation} (see Fig. 2 in \cite{maffettone2022continuification}). 

%\textcolor{blue}{For example, should we consider agent $i$ at $7/8\pi$ interacting with $j$ at $-7/8\pi$, the interaction distance to be considered is $-1/4\pi$}.

%\subsection{Continuified macroscopic dynamics}
Assuming the number of agents to be sufficiently large, the macroscopic collective dynamics can be adequately approximated through the mass balance equation \cite{bernoff2011primer}
\begin{equation}
    \rho_t(x,t)  + \left[\rho(x,t) V(x,t)\right]_x = q(x, t)
    \label{eq:controlled_model},
\end{equation}
% \begin{equation}
%     \rho_t(x,t)  + \left[\rho(x,t) V(x,t)\right]_x = 0, \ \forall x \in  \mathcal{S}, \ \forall t\geq 0
%     \label{eq::macro_model},
% \end{equation}
where $\rho : \mathcal{S} \times \mathbb{R}_{\geq 0} \to \mathbb{R}_{\geq 0}$ is the density profile of the agents on $\mathcal{S}$ at $t$ such that $\int_\mathcal{S} \rho(x,t) \,\mathrm{d}x=N$ for any $t$, and
$V$ is the velocity field, which can be expressed as
\begin{equation}
    V(x, t) = \int_{-\pi}^\pi f\left(\left\{x,y\right\}_\pi\right)\rho(y, t)\,\mathrm{d}y = (f * \rho) (x, t)
    \label{eq:V}.
\end{equation} 
%This function is positive and such that, when integrated over a subset of $\mathcal{S}$, it returns the number of agents in that subset. 
The function $q$ represents the macroscopic control input, which we first write as a mass source/sink to simplify derivations, but then recast as a controlled velocity field.

The boundary and initial conditions are given as follows:
\begin{align}
    \rho(-\pi,t) = \rho(\pi,t), \quad \forall t\geq 0 \label{eq::periodicity},\\
    \rho(x, 0) = \rho^0(x), \quad \forall x \in  \mathcal{S}.
\end{align}
We remark that $V$ is periodic by construction, as it comes from a circular convolution. This ensures that, in the open-loop scenario, when $q=0$, mass is conserved, that is $\mathrm{d}/\mathrm{d}t\int_{\mathcal{S}}\rho(x, t)\,\mathrm{d}x = 0$ (integrating by parts).

%\subsection{Problem statement}\label{sec:prob_stat}
Given some desired periodic smooth density profile, $\rho^\text{d}(x, t)$, associated with the target agents' configuration, and such that $\Vert \rho^\mathrm{d}\Vert_2 \leq M$ and $ \Vert \rho^\mathrm{d}_x \Vert_2\leq L$ at any $t$, the continuification control problem is that of finding the control inputs $u_i,\ i=1,2,\dots,N$ in \eqref{eq:themodel} such that
\begin{equation}
    \lim_{t\rightarrow \infty} \Vert{\rho^\text{d}(\cdot, t)}-\rho(\cdot, t)\Vert_2 =0,
    \label{eq:controlgoal}
\end{equation} 
for agents starting from any initial configuration $x_i(0)=x_{i0}, \ i=1,\ldots,N$.

To solve this problem, we first choose $q$ in \eqref{eq:controlled_model} as 
\begin{multline}\label{eq:q}
     q(x, t) = K_\mathrm{p}e(x, t) - \left[e(x, t)V^\mathrm{d}(x, t)\right]_x\\-\left[ \rho(x, t)V^\mathrm{e}(x, t)\right]_x,
\end{multline}
where $K_\mathrm{p}$ is a positive control gain, $e= \rho^\mathrm{d}-\rho$, $V^\text{d} = (f * \rho^\mathrm{d})$, $V^\text{e} = (f * e)$, and we consider the reference dynamics
\begin{equation}
    \rho^\mathrm{d}_t(x,t) + \left[\rho^\mathrm{d}(x,t)V^\mathrm{d}(x, t)\right]_x = 0
    \label{eq:ref_model},
\end{equation}
fulfilling initial and boundary conditions similar to \eqref{eq:controlled_model}. 
As shown in \cite{maffettone2022continuification}, such a choice ensures that the density $\rho$ globally asymptotically converges to $\rho^\mathrm{d}$.
% \begin{proof}
% By substituting \eqref{eq:q} into \eqref{eq:errmodel}, we get
% \begin{align}\label{eq:controlled_err}
%     e_t(x, t) = -K_\mathrm{p}e(x, t),
% \end{align}
% that, being $K_\mathrm{p}$ positive, converges towards 0 starting from any initial condition. 
% % We remark that the \eqref{eq:controlled_err} is a linear proportional integral dynamics and its transient and steady-state properties can be modulated using $K_\mathrm{p}$ and $K_\mathrm{i}$. We also remark that global asymptotic stability is achieve also when $K_\mathrm{i} = 0$, yet losing some interesting robustness properties that will be further discussed later. 
% \end{proof}

%\subsection{Discretization and microscopic control}\label{subsec:discretization}
Then, we recast the macroscopic controlled model \eqref{eq:controlled_model} to include $q$ as a control action on the velocity field, that is, 
\begin{align}\label{eq:macro_U_control}
    \rho_t(x,t)  + \left[\rho(x,t) (V(x,t) + U(x, t))\right]_x = 0,
\end{align}
where $U$ is an auxiliary function computed from the linear PDE 
\begin{align}
    \left[\rho(x, t) U(x, t)\right]_x = -q(x, t).
    \label{eq:v}
\end{align}
% The case $\rho= 0$ corresponds to a case in which $q(x,t)$ is effectively behaving as a source/sink, changing the mass of the system (impossible without affecting the total number of agents in the system).
Integrating \eqref{eq:v}, we obtain (assuming $\rho\ne 0$)
\begin{align}\label{eq:U}
    U(x, t) = -\frac{1}{\rho(x, t)}\left[\int q(y, t)\,\mathrm{d}y + q(-\pi, t)\right].
\end{align}

Finally, we compute the velocity input acting on agent $i$ at the microscopic level by spatially sampling $U$ at $x_i$, that is,
\begin{align}
    u_i(t) = U(x_i, t), \quad i=1,2,\ldots N.
    \label{eq:u_i}
\end{align}

The main limitation of this approach is the non-local nature of the control action. Since \eqref{eq:q} is based on the convolution $V^\mathrm{e}$, agent $i$ must have global knowledge of $e$ to compute $u_i$. Moreover, as the choice of $q$ is based on some cancellations of the macroscopic dynamics of the system, the robustness to structural perturbations needs to be properly assessed. In this study, we address both of these issues.

%\begin{remark}
% We remark the following. (i) In a decentralised scenario, each agent needs to posses, exchange, or estimate enough information about the others to compute $U(x_i, t)$, which is non-local, as it is based on $q$, that is defined via some circular convolutions. (ii) The assumption that $\rho$ is nonzero is reasonable, as agents will estimate the density from their own positions, and hence we can choose a smoothing kernel such that $U$ is always well-defined. %Here, we use a Gaussian kernel estimation, adapted to take into account the domain's periodicity.
% Moreover, since we are interested in discretizing the spatial control action, we know that $\rho$ will be different from zero at least where there are effectively agents to control, that is, $U$ is surely well-defined where we need to discretize it. (iii) The discretized controller will fulfill asymptotic convergence of agents' density to the desired one only in the limit of infinite number of agents. 

\section{Limited sensing capabilities}\label{sec:lim_sens}
%\begin{figure}
%   \centering
%    \includegraphics[width=0.35\textwidth]{figures/limited_sensing/no_integral_action_lim_sens.eps}
 %        \caption{Phase portrait of the system bounding $(\Vert e \Vert_2^2)t$, in the case of limited sensing capabilities.}
 %        \label{fig::limsens_no_int}
%\end{figure}
% The main limitation of the strategy in \cite{maffettone2022continuification} is the non-local nature of the control action. As \eqref{eq:q} is based on the convolution $V^\mathrm{e}$, agent $i$ needs to have a global knowledge of $e$ to compute $u_i$.
To relax the assumption of unlimited sensing, we assume agents can only sense an interval $[-\Delta,\Delta]$, with $\Delta>0$ centered at their position. %Note that setting $\Delta = \pi$ corresponds to agents having global knowledge of the entire domain $[-\pi,\pi]$.
The macroscopic control action in \eqref{eq:q} becomes
\begin{multline}
    \hat{q}(x, t)= K_\mathrm{p}e(x, t) - \left[e(x, t)V^\mathrm{d}(x, t)\right]_x \\-\left[ \rho(x, t)\hat{V}^\mathrm{e}(x, t)\right]_x,
\end{multline}
where $\hat{V}^\mathrm{e} = (\hat{f}*e)$, and
$\hat{f}$ is a modified velocity interaction kernel defined as
\begin{align}
    \hat{f}(z) = f(z) \Pi(z, \Delta),
\end{align}
with $\Pi(z, \Delta)$ being a rectangular window of size $2\Delta$, centered at the origin. 
%Specifically,
% \begin{align}
%     \Pi(z, \Delta) = \begin{cases}
%         1 \;\;&\mathrm{if} \;\; \vert z\vert \leq \Delta,\\
%         0 &\mathrm{otherwise}.
%     \end{cases}
% \end{align}
%Notice that we consider limited sensing capabilities only affects $V^\mathrm{e}$  and not $V^\mathrm{d}$, as we assume agents are aware of the desired density to achieve. 

Using $\hat{q}$ instead of $q$ as input to the macroscopic model \eqref{eq:controlled_model} yields the following error dynamics:
\begin{multline}\label{eq:err_lim_sens}
    e_t(x, t) = -K_\mathrm{p}e(x, t)  +\left[\rho^\mathrm{d}(x, t) \Tilde{V}(x, t)\right]_x \\-\left[e(x, t) \Tilde{V}(x, t)\right]_x,
\end{multline}
where $\tilde{V} = (g*e)$ and $g := \hat{f}-f$.

\begin{theorem} [LAS with limited sensing capabilities]\label{th:lim_sens}
    % The origin is a locally asymptotically stable equilibrium point of the error dynamics  \eqref{eq:err_lim_sens}.
    The error dynamics \eqref{eq:err_lim_sens} locally asymptotically converges to 0.
\end{theorem}
\begin{proof}
Choosing $\Vert e\Vert_2^2$ as a candidate Lyapunov function for \eqref{eq:err_lim_sens}, we get (omitting dependencies for simplicity)
% \begin{multline}\label{eq:norm_err_dyn_lim_sens}
%     (\Vert e\Vert_2^2)_t = -2K_\mathrm{p} \Vert e\Vert_2^2 + 2\int_{\mathcal{S}}e(x, t)\left[\rho^\mathrm{d}(x, t)\Tilde{V}(x, t)\right]_x \,\mathrm{d}x \\- 2\int_{\mathcal{S}}e(x, t)\left[e(x, t)\Tilde{V}(x, t)\right]_x \,\mathrm{d}x.
% \end{multline}
% Equation \eqref{eq:norm_err_dyn_lim_sens} can be rewritten as (omitting explicit dependencies for simplicity)
% \begin{multline}
%     (\Vert e\Vert_2^2)_t = -2K_\mathrm{p} \Vert e\Vert_2^2 -2K_\mathrm{i} \int_0^t \Vert e\Vert_2^2\,\mathrm{d}\tau \\+ 2\int_\mathcal{S} (e\rho^\mathrm{d}_x\Tilde{V} + e\rho^\mathrm{d}\Tilde{V}_x) \,\mathrm{d}x + 2 \int_\mathcal{S}e_x e\Tilde{V}\,\mathrm{d}x + \cancel{\left[e^2\Tilde{V}\right]_{-\pi}^\pi},
% \end{multline}
% where we applied the product derivative and integration by parts (the constant term is 0 because $e$ and $\Tilde{V}$ are periodic by construction). For an interpretation of the derivative of a convolution, see Lemma \ref{lem:young_deriv}. We can further write
% \begin{multline}
%     (\Vert e\Vert_2^2)_t = -2K_\mathrm{p} \Vert e\Vert_2^2 -2K_\mathrm{i} \int_0^t \Vert e\Vert_2^2\,\mathrm{d}\tau \\+ 2\int_\mathcal{S} (e\rho^\mathrm{d}_x\Tilde{V} + e\rho^\mathrm{d}\Tilde{V}_x) \,\mathrm{d}x + \int_\mathcal{S}(e^2)_x\Tilde{V}\,\mathrm{d}x,
% \end{multline}
% that is, applying again integration by parts and exploiting the periodic nature of the involved functions
\begin{multline}
    (\Vert e\Vert_2^2)_t = 2\int_{\mathcal{S}}e e_t \,\mathrm{d}x= -2K_\mathrm{p} \Vert e\Vert_2^2  - \int_\mathcal{S}e^2\Tilde{V}_x\,\mathrm{d}x\\+ 2\int_\mathcal{S} (e\rho^\mathrm{d}_x\Tilde{V} + e\rho^\mathrm{d}\Tilde{V}_x) \,\mathrm{d}x, \label{eq:error}
\end{multline}
where we computed product derivatives and used integration by parts taking into account the periodicity of the functions. 
\begin{figure}
   \centering
\includegraphics[width=0.35\textwidth]{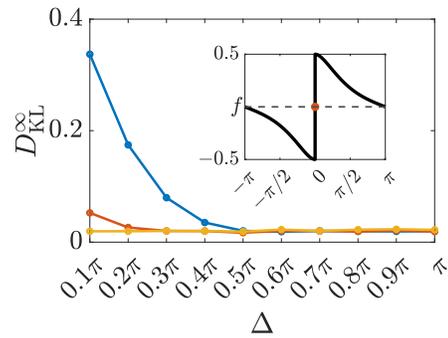}
         \caption{KL divergence at the end of a monomodal regulation trial, for different values of $\Delta$ and $K_\mathrm{p}$ ($K_\mathrm{p}= 10$ in blue, $K_\mathrm{p}= 100$ in orange and $K_\mathrm{p}= 1000$ in yellow). In the inset, we show the repulsive interaction kernel.}
         \label{fig::limsens_kl}
\end{figure}
\begin{figure}
     \centering
     \begin{subfigure}[b]{0.23\textwidth}
         \centering
         \includegraphics[width=\textwidth]{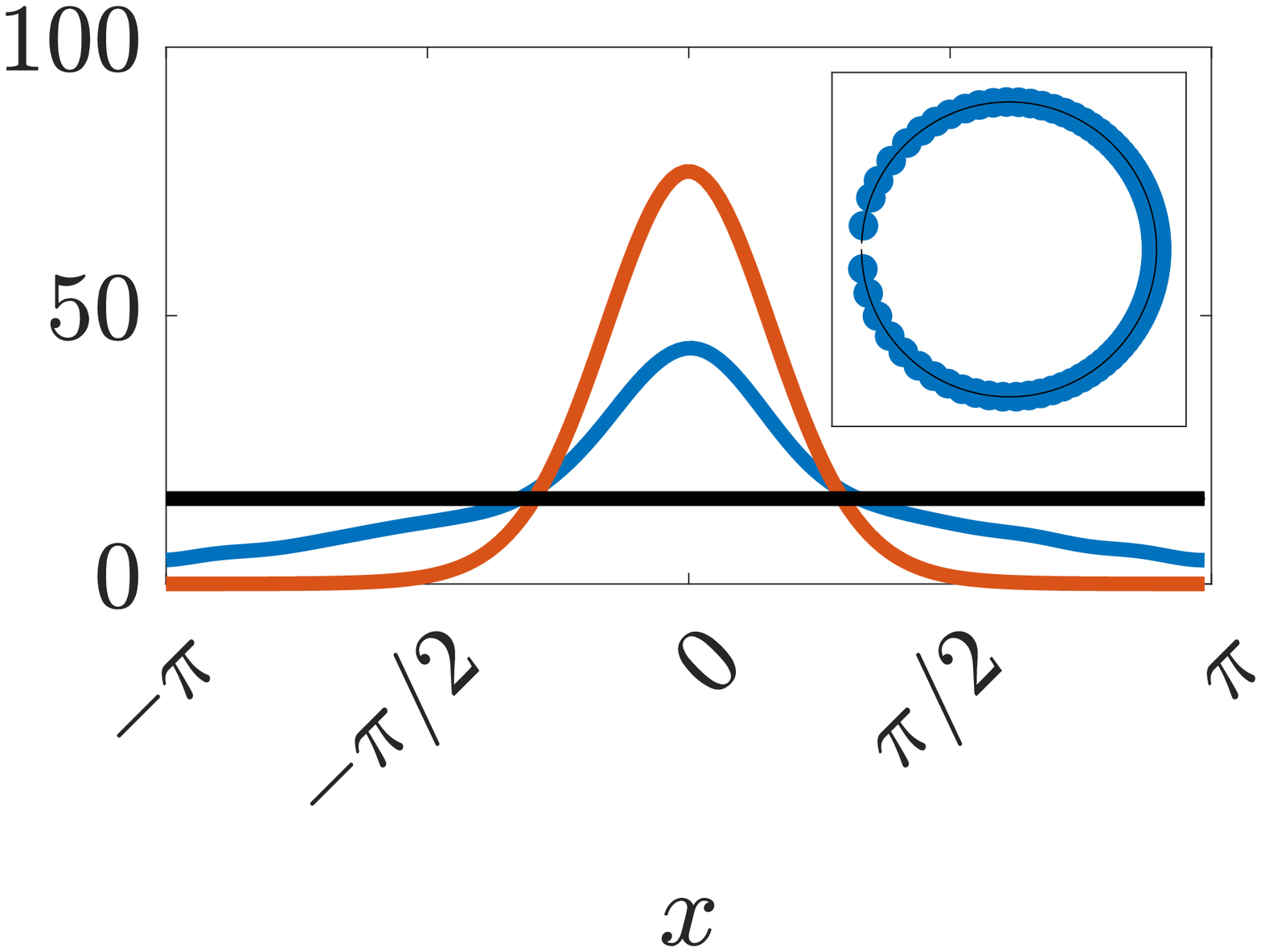}
         \caption{$\Delta = 0.1\pi$}
         \label{subfig::lim_sens_01}
     \end{subfigure}
     \begin{subfigure}[b]{0.23\textwidth}
         \centering
         \includegraphics[width=\textwidth]{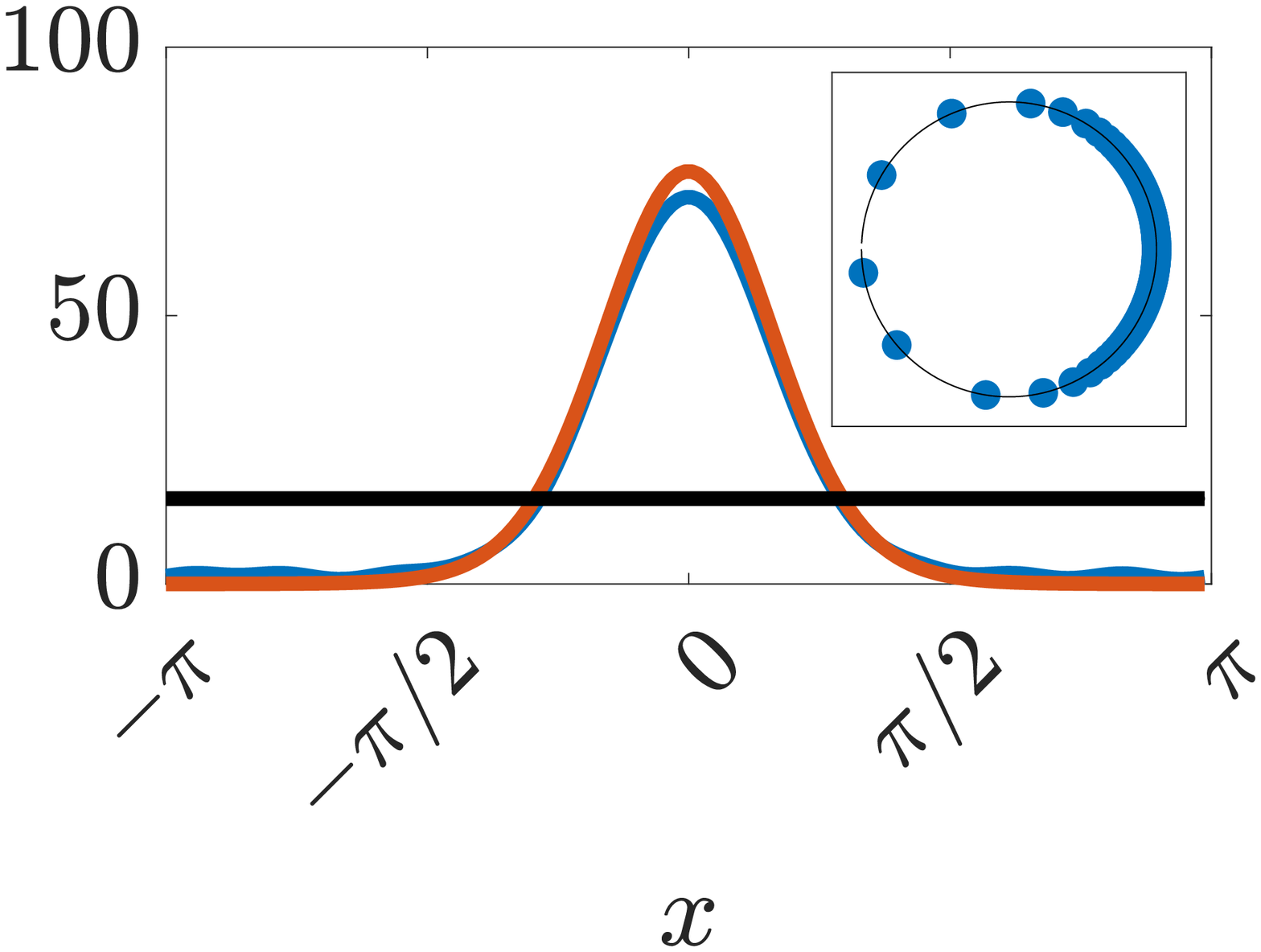}
         \caption{$\Delta = 0.4\pi$}
         \label{subfig::lim_sens_04}
     \end{subfigure}
     \begin{subfigure}[b]{0.23\textwidth}
         \centering
         \includegraphics[width=\textwidth]{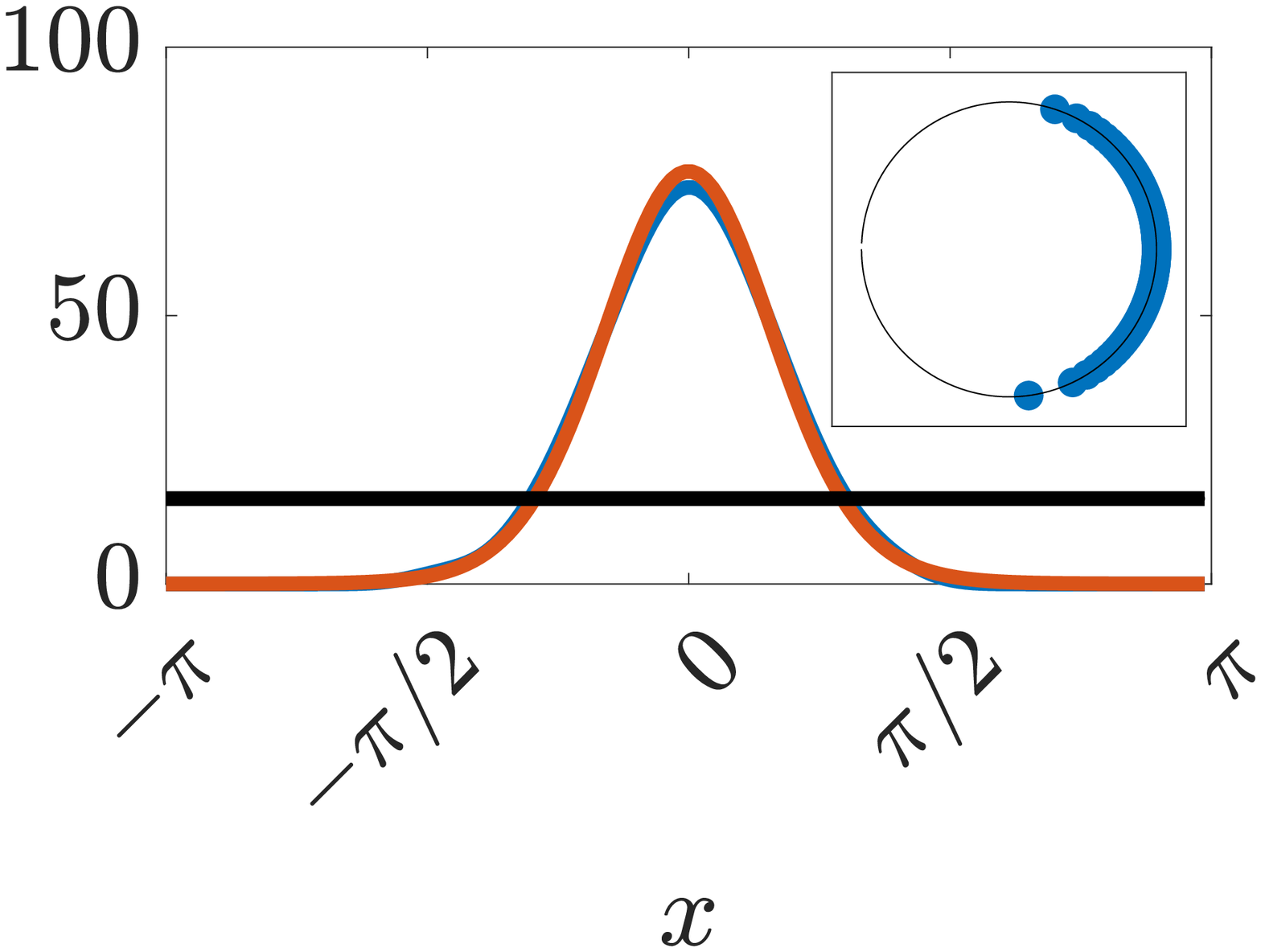}
         \caption{$\Delta = 0.7\pi$}
         \label{subfig::lim_sens_07}
     \end{subfigure}
     \begin{subfigure}[b]{0.23\textwidth}
         \centering
         \includegraphics[width=\textwidth]{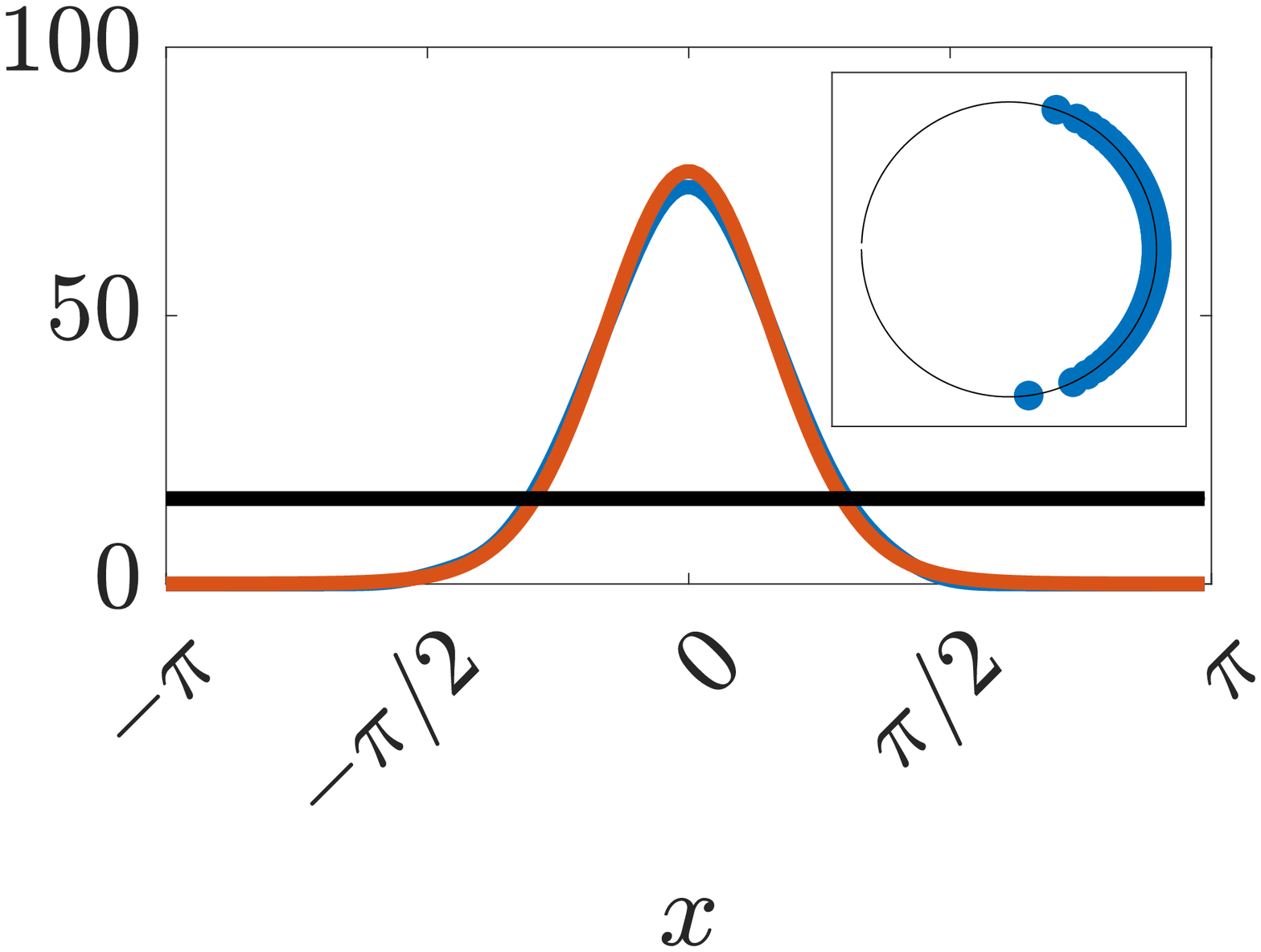}
         \caption{$\Delta = \pi$}
         \label{subfig::unlim_sens}
     \end{subfigure}
        \caption{Steady-state ($t=t_\mathrm{f}$) comparison between the agents distribution (blue line) and the desired one (orange line) when the agents are started from the initial distribution (black line) for increasing sensing abilities of the agents when $K_\mathrm{p} = 10$. Panel (d) shows the case when sensing is unlimited. In the inset of each panel, we display the discrete formation of the agents at the end of the trial.}
        \label{fig::fin_disp_lim_sens}
\end{figure}
% \begin{figure}
%    \centering
% \includegraphics[width=0.35\textwidth]{figures/limited_sensing/kp_varying_fixed_sensing.eps}
%          \caption{Time evolution of the KL divergence for different values of $K_\mathrm{p} $, assuming $\Delta = 0.1\pi$.}
%          \label{fig::fix_sens_kp_var}
% \end{figure}
Using the definition of $L^1$-norm (see Definition \ref{def:Lp_norm}),  applying Holder's inequality with $n=3$, $p_1=p_2=2$, and $p_3 = \infty$ (see Lemma \ref{th:holder}), invoking Young's convolution inequality with $r=\infty$ and $p=q=2$ (see Lemma \ref{th:young_inequality}), and recalling the assumption on the $L^2$ boundedness of $\rho^\mathrm{d}$ and $\rho^\mathrm{d}_x$ by constants $L$ and $M$, we find
\begin{multline}
    \left\vert \int_\mathcal{S}e\rho^\mathrm{d}_x\Tilde{V}\,\mathrm{d}x\right\vert \leq \int_\mathcal{S}\vert e\rho^\mathrm{d}_x\Tilde{V}\vert\,\mathrm{d}x = \Vert e\rho^\mathrm{d}_x\Tilde{V}\Vert_1 \leq\\ \leq \Vert e\Vert_2 \Vert \rho^\mathrm{d}_x\Vert_2 \Vert\Tilde{V} \Vert_\infty\leq L\Vert e\Vert_2^2\Vert g\Vert_2,
\end{multline}
\begin{multline}
    \left\vert \int_\mathcal{S}e\rho^\mathrm{d}\Tilde{V}_x\,\mathrm{d}x\right\vert \leq \int_\mathcal{S}\vert e\rho^\mathrm{d}\Tilde{V}_x\vert\,\mathrm{d}x = \Vert e\rho^\mathrm{d}\Tilde{V}_x\Vert_1 \leq\\ \leq \Vert e\Vert_2 \Vert \rho^\mathrm{d}\Vert_2 \Vert\Tilde{V}_x \Vert_\infty\leq M\Vert e\Vert_2^2\Vert g_x\Vert_2,
\end{multline}
\begin{multline}
    \left\vert \int_\mathcal{S} e^2\Tilde{V}_x\,\mathrm{d}x\right\vert \leq\int_\mathcal{S} \vert e^2\Tilde{V}_x \vert \,\mathrm{d}x = \Vert ee\Tilde{V}_x\Vert_1 \leq \\ \leq \Vert e\Vert^2_2  \Vert \Tilde{V}_x\Vert_\infty  \leq \Vert e\Vert^3_2 \Vert g_x\Vert_2.
\end{multline}

Using these bounds, from \eqref{eq:error} we can write
\begin{align}\label{eq:ineq_lim_sens}
    (\Vert e \Vert_2^2)_t \leq (-2K_\mathrm{p}+2M\Vert g_x \Vert_2+2L\Vert g \Vert_2 + \Vert g_x \Vert_2 \Vert e \Vert_2)\Vert e \Vert_2^2.
\end{align}
Then, for any $\gamma>0$, if $\Vert e\Vert_2 < \gamma$, the error asymptotically tends to the origin for $K_\mathrm{p}$ sufficiently large $\left[K_\mathrm{p}>(M+\gamma/2)\Vert g_x \Vert_2+L\Vert g \Vert_2\right]$, hence local asymptotic convergence is proved. 
% \begin{remark}
%     Local asymptotic stability is achieved also if $K_\mathrm{i}=0$.
% \end{remark}
\end{proof}
%\begin{corollary} \label{cor:limsens}
%For $K_\mathrm{p}>M\Vert g_x \Vert_2+L\Vert g \Vert_2$, the width of the region of asymptotic stability of the origin of the error system can be increased by  %trajectories $\Vert e(\cdot, t)\Vert_2^2$ are bounded by those of a system converging to 0 from an arbitrarily large set of initial conditions. 
%\end{corollary}
%\begin{proof}
%We can rewrite \eqref{eq:ineq_lim_sens} as
%\begin{align}
%    % \dot{\eta} \leq -a \eta +c \eta\sqrt{\eta}:=h(\eta),
%\end{align}
%where $\eta = \Vert e \Vert_2^2$, $a=-2K_\mathrm{p}+2M\Vert g_x \Vert_2+2L\Vert g \Vert_2$, and $c = \Vert g_x \Vert_2$. Thanks to the comparison lemma (see Lemma \ref{lemma:comparison_lemma}), we know that the trajectories of the system $v_t = h(v)$ bounds $\eta(t)$ at any $t$, if $\eta(0) \leq v(0)$. As we are assuming $K_\mathrm{p}$ is such that $a>0$, the bounding system $v_t$ exhibit a locally stable equilibrium in the origin (with basin of attraction $[0, a^2/c^2)$) and an unstable one in $a^2/c^2$, as reported in Fig. \ref{fig::limsens_no_int}. The greater $K_\mathrm{p}$, the greater $a$, and consequently the basin of attraction of the origin. %Thus $\eta(t)$ is always bounded by the trajectories $v(t)$.
%\end{proof}

\subsubsection*{Numerical validation}
We consider the same framework, control discretization and numerical set-up as in \cite{maffettone2022continuification}. In particular, we refer to a mono-modal regulation scenario, where a repulsive swarm of $N=100$ agents, starting evenly displaced in $\mathcal{S}$, is required to achieve a desired density profile given by a von Mises function, with mean $\mu = 0$ and concentration coefficient $k=4$. The pairwise interactions between agents is modelled via a repulsive Morse potential, depicted in the inset of Fig. \ref{fig::limsens_kl}, given by
\begin{align}\label{eq:Morse}
    f(x) = \mathrm{sign}(x)\left[-G \mathrm{e}^{-\vert x \vert /L} + \mathrm{e}^{-\vert x \vert}\right],
\end{align}
where the characteristic parameters, modulating the strength and characteristic distance of the attractive term,
are $G=L=0.5$, making the repulsion term dominant. 

We run several trials of duration $t_\mathrm{f} = 6$.
In each trial, we consider a different sensing radius $\Delta$, spanning from $0.1\pi$ to $\pi$. At the end of each trial, we record the steady-state Kullback-Leibler (KL) divergence between $\hat{\rho}$ and $\hat{\rho}^\mathrm{d}$ (equivalent to $\rho$ and $\rho^\mathrm{d}$, but normalized to sum to 1), $D_\mathrm{KL}^\infty$ \cite{kullback1951information}. The results of such a numerical investigation are reported in Fig. \ref{fig::limsens_kl}, for different values of $K_\mathrm{p}$.  They show that: (i) for large values of $K_\mathrm{p}$, performance is independent from the specific sensing radius that is given to the agents, and (ii) for smaller values of $K_\mathrm{p}$, a limited knowledge of the domain can still guarantee a performance level that is comparable to the case of $\Delta = \pi$. For example, when considering $K_\mathrm{p}=10$, choosing $\Delta = 0.4\pi$ makes $D_\mathrm{KL}^\infty$ comparable to the case of unlimited sensing capabilities. 
For the case $K_\mathrm{p}=10$, we also report in Fig. \ref{fig::fin_disp_lim_sens} the final configuration of the swarm (both in discrete and continuified terms) for different values of the sensing radius.
We remark that the non-zero residual $D_\mathrm{KL}$ is due to the discretization process. 

\section{Structural perturbations}\label{sec:struc_pert}
% Another limitation of the continuification control in \cite{maffettone2022continuification} is its robustness to structural perturbations, as it is based on some macroscopic cancellations of the system dynamics.
Next, we assess the robustness of the approach to two classes of perturbations, the first acting additively on the macroscopic velocity field and the second on the interaction kernel. 
% We refer to the first case as spatio-temporal disturbances, and to the second as parametric uncertainties. 

% Each analysis is complemented with a numerical simulation, considering a similar scenario to the mono-modal regulation that is discussed in \cite{maffettone2022continuification}. We consider a swarm of $N=100$ agents, whose pairwise interactions are regulated by a repulsive Morse potential (i.e., the characteristic parameters are set as $G=L=0.5$). In the absence of control, agents would spread over the domain, until reaching a constant density. We further assume they start evenly distributed over $\mathcal{S}$ (i.e., $\rho^0 = N/2\pi$), We want them to achieve desired density profile that is a von Mises function with mean $\mu =0$ and concentration coefficient $k = 4$.

\subsection{Spatio-temporal perturbations of the velocity field}\label{subsec:dist}
We assume that perturbations of the microscopic dynamics can be captured at the macroscopic level by means of some spatio-temporal velocity field $d(x, t)$ affecting \eqref{eq:controlled_model}. The macroscopic controlled model becomes
\begin{align}\label{eq:macro_const_dist}
    \rho_t(x, t) + \left[\rho(x, t) (V(x, t)+d(x, t))\right]_x = q(x, t), 
\end{align}
where we assume $d(-\pi, t) = d(\pi, t)$ for any $t$ and $d, d_x \in L^\infty$ at any $t$ so that that there exist two positive constants $D_1$ and $D_2$  bounding the $L^\infty$-norm of $d$ and $d_x$, respectively.

Substituting \eqref{eq:q} into \eqref{eq:macro_const_dist} and taking into account the reference dynamics \eqref{eq:ref_model} yields
\begin{multline}\label{eq:err_cons_dist}
    e_t(x, t) = -K_\mathrm{p}e(x, t) + \left[(\rho^\mathrm{d}(x, t) - e(x, t))d(x, t)\right]_x.
\end{multline}
% where we used that $\rho = \rho^\mathrm{d}-e$. 

\begin{theorem}[Bounded convergence in the presence of velocity perturbations]
$\,$ There exists a threshold value $D_2<\kappa<+\infty$ such that, if $2K_\mathrm{p}>\kappa$, the dynamics of the squared error norm is bounded and 
 $$
 \lim_{t \rightarrow \infty}{\Vert e(\cdot, t)\Vert_2}\leq \frac{2LD_1+2MD_2}{\kappa-D_2}
 $$
 Hence, the upper bound on the steady-state error can be made arbitrarily small by choosing $\kappa$ sufficiently large.
\end{theorem}
% \begin{theorem}[LAS with velocity perturbations] \label{th:dist}
% The origin is a locally asymptotically stable equilibrium of the error system \eqref{eq:err_cons_dist}.
% \end{theorem}
\begin{proof}
%Let us assume $\Vert e\Vert_2^2$ is a candidate Lyapunov function for \eqref{eq:err_cons_dist}. 
Taking into account \eqref{eq:err_cons_dist}, we write the dynamics of  $\Vert e\Vert_2^2$ (omitting dependencies for simplicity) as
\begin{multline}\label{eq:err_norm_dyn_cons_dist1}
    (\Vert e\Vert_2^2)_t = 2\int_\mathcal{S}e e_t\,\mathrm{d}x =-2K_\mathrm{p}\Vert e\Vert_2^2 - \int_\mathcal{S}e^2d_x\,\mathrm{d}x \\+ 2\int_\mathcal{S}(e\rho_x^\mathrm{d}d + e\rho^\mathrm{d}d_x) \,\mathrm{d}x,
        % (\Vert e\Vert_2^2)_t = -2K_\mathrm{p}\Vert e\Vert_2^2 - 2 K_\mathrm{i}\int_0^t \Vert e(\cdot, t)\Vert_2^2 \,\mathrm{d}\tau + 2d\int_\mathcal{S}e(x, t)\rho_x^\mathrm{d}(x, t)\,\mathrm{d}x - 2d\int_\mathcal{S}e(x, t)e_x(x, t)\,\mathrm{d}x=\\=-2K_\mathrm{p}\Vert e\Vert_2^2 - 2 K_\mathrm{i}\int_0^t \Vert e(\cdot, t)\Vert_2^2 \,\mathrm{d}\tau + 2d\int_\mathcal{S}e(x, t)\rho_x^\mathrm{d}(x, t)\,\mathrm{d}x - d \int_\mathcal{S} (e^2(x, t))_x \,\mathrm{d}x=\\=-2K_\mathrm{p}\Vert e\Vert_2^2 - 2 K_\mathrm{i}\int_0^t \Vert e(\cdot, t)\Vert_2^2 \,\mathrm{d}\tau + 2d\int_\mathcal{S}e(x, t)\rho_x^\mathrm{d}(x, t)\,\mathrm{d}x,
\end{multline}
where we computed product derivatives and applied integration by parts exploiting the periodicity of the functions.
Similar to the proof of Theorem \ref{th:lim_sens}, we apply Definition \ref{def:Lp_norm}, Holder's inequality with $n=3$, $p_1 = p_2 = 2$ and $p_3 = \infty$ (see Lemma \ref{th:holder}), and exploit the bounds on $\rho^\mathrm{d}$, $\rho^\mathrm{d}_x$, $d$ and $d_x$, to derive the following inequalities for the terms in \eqref{eq:err_norm_dyn_cons_dist1}:
\begin{multline}
    \left\vert \int_\mathcal{S} e \rho^\mathrm{d}_x d\,\mathrm{d}x \right\vert \leq  \int_\mathcal{S} \vert e \rho^\mathrm{d}_x d \vert\,\mathrm{d}x = \Vert e\rho^\mathrm{d}_x d\Vert_1\leq \\ \leq \Vert e\Vert_2 \Vert \rho^\mathrm{d}_x\Vert_2 \Vert d\Vert_\infty \leq  L D_1 \Vert e\Vert_2,
\end{multline}
\begin{multline}
    \left\vert \int_\mathcal{S} e \rho^\mathrm{d} d_x\,\mathrm{d}x \right\vert \leq  \int_\mathcal{S} \vert e \rho^\mathrm{d} d_x \vert\,\mathrm{d}x = \Vert e\rho^\mathrm{d} d_x\Vert_1\leq \\ \leq \Vert e\Vert_2 \Vert \rho^\mathrm{d}\Vert_2 \Vert d_x\Vert_\infty \leq  M D_2\Vert e\Vert_2,
\end{multline}
\begin{multline}
    \left\vert \int_\mathcal{S} e^2 d_x\,\mathrm{d}x \right\vert \leq  \int_\mathcal{S} \vert e^2 d_x \vert\,\mathrm{d}x = \Vert ee d_x\Vert_1\leq \\ \leq \Vert e\Vert_2^2 \Vert d_x\Vert_\infty \leq  D_2\Vert e\Vert_2^2.
\end{multline}

Hence, we obtain
\begin{multline}\label{eq:diff:ineq_const_dist}
        (\Vert e \Vert_2^2)_t \leq \left(-2K_\mathrm{p} +D_2\right)\Vert e \Vert_2^2 + \left(2LD_1+2MD_2\right)\Vert e\Vert_2.
\end{multline}
% We can conclude that, for any $\gamma>0$, if $\Vert e\Vert_2 <\gamma$, the error will approach zero asymptotically, for $K_\mathrm{p}$ sufficiently large $\left[K_\mathrm{p}>D_2/2 +LD_1/\gamma + MD_2/\gamma\right]$.
% \begin{remark}
% If $K_i = 0$ (no integral action implemented in $q$),  then $b=0$, and the following differential inequality can be studied:
% \begin{align}
%     \eta_t \leq -a\eta+c\sqrt{\eta} := h(\eta).
% \end{align}
% The bounding dynamical system described by $v_t = h(v)$ has two equlibria, an unstable one $\eta_1 = 0$ and a stable one $\eta_2 = c^2/a^2$. Increasing the proportional action embedded in $a$, $\eta_2$ gets closer and closer to the origin, yet not guaranteeing total rejection of the constant disturbance (see Fig.  \ref{subfig::const_dist_no_int}, for a graphical interpretation). 
% \end{remark}
%\end{proof}
\begin{figure}
    \centering
         \includegraphics[width=0.35\textwidth]{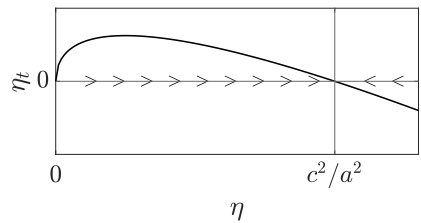}
         \caption{Phase portrait of the system bounding $\Vert e\Vert_2^2$ in the presence of spatio-temporal disturbances.}
         \label{fig::const_dist_no_int}
\end{figure}
\begin{figure}
   \centering
    \includegraphics[width=0.35\textwidth]{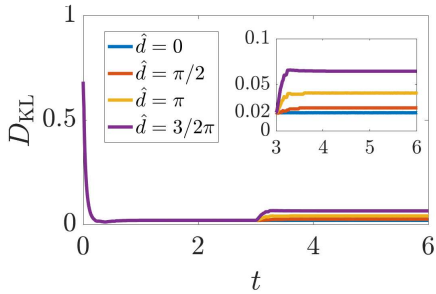}
         \caption{Time evolution of the KL divergence when a constant disturbance of amplitude $\hat{d}$ switches on at $t = 3$. In the inset, a zoom of the second half of the trial.}
         \label{fig::dist_kl}
\end{figure}
% \begin{corollary} \label{cor:dist}
% For $K_\mathrm{p}>D_2/2$, the trajectories $\Vert e(\cdot, t)\Vert_2^2$ are always bounded by those of a system asymptotically converging arbitrarily close to the origin.
% \end{corollary}
%\begin{proof}
For convenience, we rewrite \eqref{eq:diff:ineq_const_dist} as
\begin{align}\label{eq:eta_const_dist}
        \eta_t \leq -a \eta + c\sqrt{\eta}:=h(\eta),
\end{align}
where  $\eta=\Vert e\Vert_2^2$, $a = 2 K_\mathrm{p} - D_2$, and $c = 2LD_1 + 2MD_2$. Under the assumption that $2K_\mathrm{p} > D_2$ ($a>0$), the phase portrait of the bounding field $h$
yield an asymptotically stable equilibrium at (see Fig. \ref{fig::const_dist_no_int}) 
$$\frac{c^2}{a^2}=\frac{(2LD_1+2MD_2)^2}{(2K_\mathrm{p}-D_2)^2}
$$
(with basin of attraction $\mathbb{R}_{> 0}$). Thus, using Lemma \ref{lemma:comparison_lemma}, $\eta$ is bounded by $c^2/a^2$. Moreover, choosing $2K_\mathrm{p}>\kappa>D_2$, the  stable equilibrium of the bounding field $h$ can be moved arbitrarily closer to the origin, thereby proving our claim. 
\end{proof}

% Corollary \ref{cor:dist} gives information regarding the case $K_\mathrm{p}$ is not large enough to ensure local stability. In such a scenario, we know that $\Vert e \Vert_2^2$ will converge towards a bounded value.
\subsubsection*{Numerical validation}
We consider the same scenario presented in the previous section but assuming that a disturbance $d(x, t) = \hat{d}w(t)$, where $\hat{d}$ is a constant and $w(t) = \mathrm{step}(t-t_\mathrm{f}/2)$, is acting on the macroscopic dynamics. Setting $K_\mathrm{p} = 10$ and considering different values of $\hat{d}$, we obtain the results reported in Fig. \ref{fig::dist_kl}. 
As expected, in the presence of the disturbances, the KL divergence remains bounded and decreases as the control gain
$K_\mathrm{p}$ increases. For example, the steady-state value of the KL divergence decreases from 0.06 when $K_\mathrm{p}=10$ to less than 0.02 when $K_\mathrm{p}\geq 100$.
%\begin{table}
%    \centering
%    \begin{tabular}{|c|c|c|c|}
%    \hline
%         $K_\mathrm{p}$ & 10 & 100 & 1000 \\\hline
%         $D_\mathrm{KL}(t_\mathrm{f})$ & 0.06 & 0.02 & 0.02
%         \\\hline
%    \end{tabular}
%    \caption{Steady state KL divergence, in the presence of the disturbance with $\hat{d}=3/2 \pi$, for different values of $K_\mathrm{p}$.}
%    \label{tab:dist}
%\end{table}

% In accordance to Corollary \ref{cor:dist}, we see that, when the disturbance comes in, the KL divergence reaches a larger, but yet acceptable, constant steady-state value. Such a value con be reduced by increasing $K_\mathrm{p}$.

\subsection{Interaction kernel perturbation}
Next, we consider the case where  structural perturbations affect the interaction kernel. We assume that the interaction kernel, $\Tilde{f}$, used to compute the macroscopic control action is different from the actual interaction kernel, $f$, influencing the agents' motion. We compute the control input as 
\begin{multline}\label{eq:q_tilde}
    \Tilde{q}(x, t)= K_\mathrm{p}e(x, t)  - \left[e(x, t)\Tilde{V}^\mathrm{d}(x, t)\right]_x \\-\left[ \rho(x, t)\Tilde{V}^\mathrm{e}(x, t)\right]_x,
\end{multline}
where $\Tilde{V}^\mathrm{d} = (\Tilde{f}*\rho^\mathrm{d})$ and $\Tilde{V}^\mathrm{e}= (\Tilde{f}*e)$.

Substituting \eqref{eq:q_tilde} %instead of \eqref{eq:q} 
into \eqref{eq:controlled_model} and considering the reference dynamics \eqref{eq:ref_model}, the error dynamics becomes
\begin{multline}\label{eq:param_unc_err}
    e_t(x, t) = -K_\mathrm{p}e(x, t)+ \left[e(x, t) \tilde{U}^\mathrm{d}(x, t)\right]_x  \\+ \left[\rho^\mathrm{d}(x, t) \tilde{U}^\mathrm{e}(x, t)\right]_x - \left[e(x, t) \tilde{U}^\mathrm{e}(x, t)\right]_x,
\end{multline}
where, letting $\tilde{g}=\Tilde{f}-f$ be the mismatch between the interaction kernels, we have 
\begin{multline}
    \Tilde{U}^\mathrm{d}(x, t) = \Tilde{V}^\mathrm{d}(x, t) - V^\mathrm{d}(x, t) = (\tilde{g}*e)(x, t),
\end{multline}
\begin{multline}
    \Tilde{U}^\mathrm{e}(x, t) = \Tilde{V}^\mathrm{e}(x, t) + V^\mathrm{e}(x, t) = (\tilde{g}*e)(x, t).
\end{multline}

\begin{theorem}[LAS with kernel perturbation]
    $\,$The error dynamics \eqref{eq:param_unc_err} locally asymptotically converges to 0, if $\tilde{g}, \tilde{g}_x \in L^2$.
\end{theorem}
\begin{proof}
Assuming $\Vert e \Vert_2^2$ to be a candidate Lyapunov function for \eqref{eq:param_unc_err}, we get
% \begin{multline}
%     (\Vert e\Vert_2^2)_t = -2K_\mathrm{p} \Vert e\Vert_2^2 -2K_\mathrm{i} \int_0^t \Vert e\Vert_2^2\,\mathrm{d}\tau + 2\int_\mathcal{S} e(\rho^\mathrm{d}\Tilde{U}^\mathrm{e})_x\,\mathrm{d}x - 2\int_\mathcal{S} e(e\Tilde{U}^\mathrm{e})_x\,\mathrm{d}+ 2\int_\mathcal{S} e(e\Tilde{U}^\mathrm{d})_x\,\mathrm{d}x
% \end{multline}
% \begin{align}
%     (\Vert e\Vert_2^2)_t = -2K_\mathrm{p} \Vert e\Vert_2^2 -2K_\mathrm{i} \int_0^t \Vert e\Vert_2^2\,\mathrm{d}\tau + 2\int_\mathcal{S} (e\rho^\mathrm{d}_x\Tilde{U}^\mathrm{e} - e\rho^\mathrm{d}\Tilde{U}^\mathrm{e}_x)
%     \,\mathrm{d}x + 2\int_\mathcal{S} e_x e\Tilde{U}^\mathrm{e}\,\mathrm{d}- 2\int_\mathcal{S} ee_x\Tilde{U}^\mathrm{d}\,\mathrm{d}x
% \end{align}
% \begin{align}
%     (\Vert e\Vert_2^2)_t = -2K_\mathrm{p} \Vert e\Vert_2^2 -2K_\mathrm{i} \int_0^t \Vert e\Vert_2^2\,\mathrm{d}\tau - 2\int_\mathcal{S} (e\rho^\mathrm{d}_x\Tilde{U}^\mathrm{e} + e\rho^\mathrm{d}\Tilde{U}^\mathrm{e}_x)
%     \,\mathrm{d}x + \int_\mathcal{S} (e^2)_x\Tilde{U}^\mathrm{e}\,\mathrm{d}- \int_\mathcal{S} (e^2)_x\Tilde{U}^\mathrm{d}\,\mathrm{d}x
% \end{align}
\begin{multline}
\label{eq:error2}
    (\Vert e\Vert_2^2)_t = \int_\mathcal{S} ee_t \,\mathrm{d}x =  -2K_\mathrm{p} \Vert e\Vert_2^2- \int_\mathcal{S}e^2\Tilde{U}^\mathrm{e}_x\,\mathrm{d}x \\+ \int_\mathcal{S} e^2 \Tilde{U}^\mathrm{d}_x\,\mathrm{d}x - 2\int_\mathcal{S} (e\rho^\mathrm{d}_x\Tilde{U}^\mathrm{e} + e\rho^\mathrm{d}\Tilde{U}^\mathrm{e}_x) \,\mathrm{d}x.
\end{multline}
where we computed the product derivatives and used integration by parts by exploiting the fact that $\Tilde{U}^\mathrm{d}$ and $\Tilde{U}^\mathrm{e}$ are periodic by construction (they come from a circular convolution). 
Using similar arguments to those above, we can establish upper bounds for the terms in \eqref{eq:error2} as follows:
\begin{multline}
    \left\vert \int_\mathcal{S}e\rho^\mathrm{d}_x\Tilde{U}^\mathrm{e}\,\mathrm{d}x\right\vert \leq \int_\mathcal{S}\vert e\rho^\mathrm{d}_x\Tilde{U}^\mathrm{e}\vert\,\mathrm{d}x = \Vert e\rho^\mathrm{d}_x\Tilde{U}^\mathrm{e}\Vert_1 \leq \\ \leq \Vert e\Vert_2 \Vert \rho^\mathrm{d}_x\Vert_2 \Vert\Tilde{U}^\mathrm{e} \Vert_\infty\leq L\Vert \tilde{g}\Vert_2\Vert e\Vert_2^2,
\end{multline}
\begin{multline}
    \left\vert \int_\mathcal{S}e\rho^\mathrm{d}\Tilde{U}^\mathrm{e}_x\,\mathrm{d}x\right\vert \leq \int_\mathcal{S}\vert e\rho^\mathrm{d}\Tilde{U}^\mathrm{e}_x\vert\,\mathrm{d}x = \Vert e\rho^\mathrm{d}\Tilde{U}^\mathrm{e}_x\Vert_1 \leq \\ \leq \Vert e\Vert_2 \Vert \rho^\mathrm{d}\Vert_2 \Vert\Tilde{U}^\mathrm{e}_x \Vert_\infty\leq M\Vert \tilde{g}_x\Vert_2\Vert e\Vert_2^2,
\end{multline}
\begin{multline}
    \left\vert \int_\mathcal{S} e^2\Tilde{U}^\mathrm{e}_x\,\mathrm{d}x\right\vert \leq\int_\mathcal{S} \vert e^2\Tilde{U}^\mathrm{e}_x \vert \,\mathrm{d}x = \Vert ee\Tilde{U}^\mathrm{e}_x\Vert_1 \leq \\ \leq\Vert e\Vert^2_2  \Vert \Tilde{U}^\mathrm{e}_x\Vert_\infty  \leq  \Vert \tilde{g}_x\Vert_2\Vert e\Vert^3_2,
\end{multline}
\begin{multline}
    \left\vert \int_\mathcal{S} e^2\Tilde{U}^\mathrm{d}_x\,\mathrm{d}x\right\vert \leq\int_\mathcal{S} \vert e^2\Tilde{U}^\mathrm{d}_x \vert \,\mathrm{d}x = \Vert ee\Tilde{U}^\mathrm{d}_x\Vert_1 \leq \\ \leq \Vert e\Vert^2_2  \Vert \Tilde{U}^\mathrm{d}_x\Vert_\infty  \leq \Vert e\Vert^2_2 \Vert \Vert \rho^\mathrm{d}\Vert_2\Vert\Tilde{g}_x\Vert_2 \leq M\Vert \tilde{g}_x\Vert_2\Vert e\Vert^2_2.
\end{multline}
Using these bounds in \eqref{eq:error2} we obtain
\begin{align}\label{eq:diff_ineq_param}
    (\Vert e \Vert_2^2)_t \leq (-2K_\mathrm{p}+3M\Vert \tilde{g}_x \Vert_2+2L\Vert \tilde{g} \Vert_2 +\Vert \tilde{g}_x \Vert_2 \Vert e \Vert_2)\Vert e \Vert_2^2.
\end{align}
Then, for any $\gamma > 0$, if $\Vert e\Vert_2 < \gamma$, the error converges to 0, for $K_\mathrm{p}$ sufficiently large, namely, $K_\mathrm{p}> \Vert\Tilde{g}_x \Vert_2\gamma/2 +3M\Vert\Tilde{g}_x\Vert_2/2+ L\Vert \tilde{g} \Vert_2$.
\end{proof}
\begin{figure}
    \centering
         \includegraphics[width=0.35\textwidth]{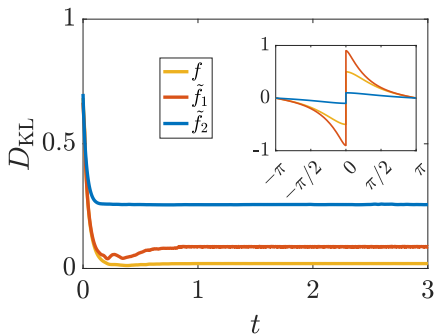}
         \caption{Time evolution of the KL divergence when the perturbed kernels $\Tilde f_1$ and $\Tilde f_2$ shown in the inset are used to compute the control action, instead of the nominal kernel $f$.}
         \label{fig::param_unc}
\end{figure}
% \begin{figure}
%     \centering
%          \includegraphics[width=0.35\textwidth]{figures/param_unc/param_unc_kp_varying.eps}
%          \caption{Time evolution of the KL divergence for different values of $K_\mathrm{p}$, when using $\Tilde{f}_2$ for computing $q$.}
%          \label{fig::param_unc_kp_varying}
% \end{figure}
% \begin{corollary}
%     Choosing $K_\mathrm{p} > (3M\Vert \tilde{g}_x \Vert_2+2L\Vert \tilde{g} \Vert_2)/2$, the trajectories $\Vert e(\cdot, t) \Vert_2^2$ are always bounded by those of a system asymptotically converging to the origin from an arbitrarily large set of initial conditions.
% \end{corollary}
% \begin{proof}
%     Equation \eqref{eq:diff_ineq_param} is analogous to \eqref{eq:ineq_lim_sens}. Then, the same considerations of Corollary \ref{cor:limsens} holds and the same argument
%     about the basin of attraction of the bounding system in the case of limited sensing capabilities (see Fig. \ref{fig::limsens_no_int}) can be done.
% \end{proof}

\subsubsection*{Numerical validation} 
We consider again the scenario used in Section \ref{sec:lim_sens} but we assume that a perturbed kernel is used to compute the macroscopic control action $q$.  %Without any loss of generality, we did the assumption of perturbing the interaction kernels without modifying their overall behavior. In particular, we chose $\Tilde{f}_1$ and $\Tilde{f}_2$ to be repulsive as $f$. 
Specifically, we test robustness against the two perturbed kernels $\Tilde{f}_1$ and $\Tilde{f}_2$, obtained by setting the characteristic parameters in \eqref{eq:Morse} to $G=L=0.1$ and $G=L=0.9$, respectively.

%, in the kernel whose nominal values were set at $G=L=0.5$.
Setting $K_\mathrm{p}= 10$, we obtain results shown in Fig. \ref{fig::param_unc} where we observe an increase in the steady-state mismatch between the distribution of the agents and the desired density as the kernel becomes more different than the nominal one ($\Tilde{f}_2$ being the worst case). As expected from the analysis, our numerical results (omitted here of brevity) confirm that the steady-state mismatch decreases as $K_\mathrm{p}$ increases approaching the nominal case in \cite{maffettone2022continuification}.  

\section{Adding a macroscopic Integral action} \label{sec:integral}
In all the cases examined earlier, some bounded mismatch between the desired and steady-state distribution of the agents remains in the presence of perturbations. To resolve this issue, we explored the inclusion of an integral action in the form
$
 K_\mathrm{i}\int_0^\tau e(x, \tau)\,\mathrm{d}\tau,
$
% in \eqref{eq:q}, modifying the macroscopic control  as
% \begin{multline}
%     q_\mathrm{I}(x, t) = K_\mathrm{p}e(x, t) + K_\mathrm{i}\int_0^\tau e(x, \tau)\,\mathrm{d}\tau \\- \left[e(x, t)V^\mathrm{d}(x, t)\right]_x-\left[ \rho(x, t)V^\mathrm{e}(x, t)\right]_x,
% \end{multline}
with $K_\mathrm{i}$ being a positive control gain, to the macroscopic control law in \eqref{eq:q}.  

\begin{figure}[t]
     \centering
     \begin{subfigure}[b]{0.23\textwidth}
         \centering
         \includegraphics[width=\textwidth]{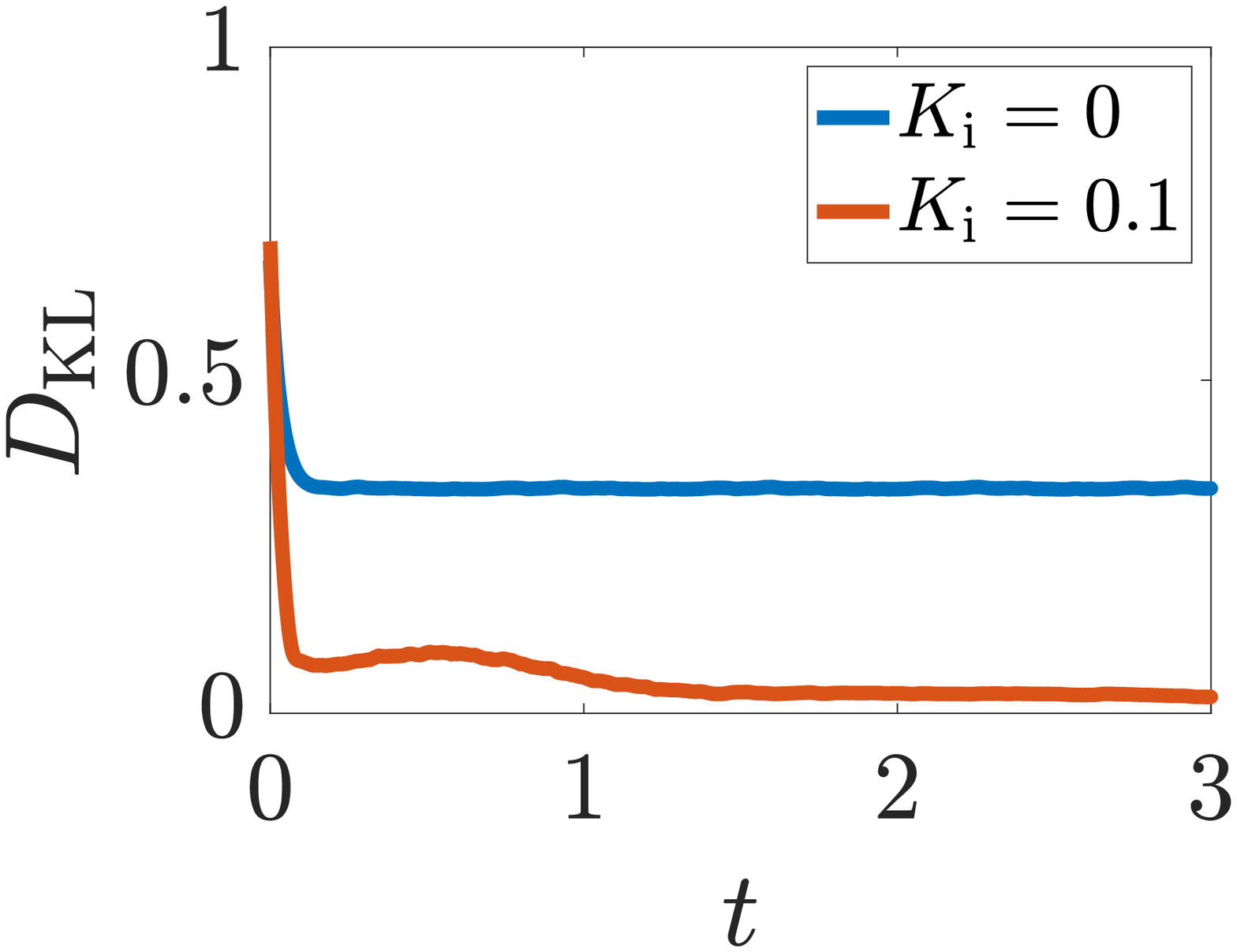}
         %\caption{$\Delta = 0.1\pi$}
         \label{subfig::lim_sens}
     \end{subfigure}
%     \begin{subfigure}[b]{0.24\textwidth}
%         \centering
%         \includegraphics[width=\textwidth]{figures/integral_action/KL_int_action_dist.eps}
  %       \caption{$\hat{d} = 3/2\,\pi$}
 %        \label{subfig::dist}
%     \end{subfigure}
     \begin{subfigure}[b]{0.23\textwidth}
         \centering
         \includegraphics[width=\textwidth]{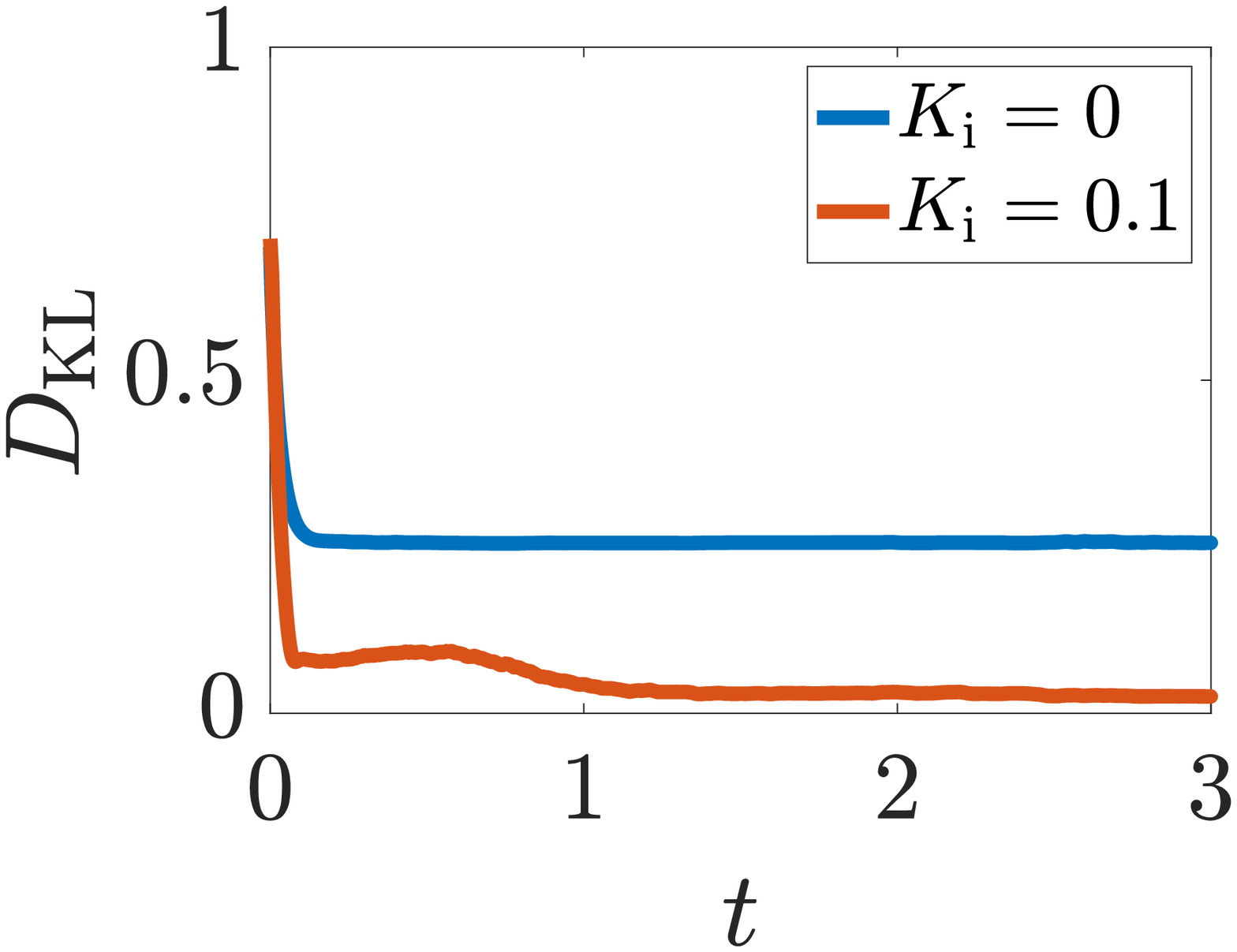}
         %\caption{$\Tilde{f}_2$}
         \label{subfig::kern_pert}
     \end{subfigure}
        \caption{Effects of a macroscopic integral action when (a) agents possess limited sensing with $\Delta=0.1\pi$ and (b) when their interaction kernel is perturbed and set equal to $\Tilde f_2$. We compare the cases of $K_\mathrm{i}=0$ and $K_\mathrm{i}=0.1$, for $K_\mathrm{p}=10$.}
        \label{fig::int_action}
\end{figure}

Preliminary results in Fig. \ref{fig::int_action} indicate a substantial improvement of the steady-state error in all the considered scenarios (for brevity only two cases are shown). These findings point at the possibility to compensate for disturbances and perturbations within a continuification-based control strategy via an additional integral action. 
The analytical characterization of the effects of the macroscopic integral action is the subject of ongoing work and will be presented elsewhere.
%Using this modified control action, where we fix $K_\mathrm{p} = 10$ and $K_\mathrm{I} = 0.1$, and considering the perturbed scenarios of the previous sections, we obtain the results in Fig. \ref{fig::int_action}, where a comparison with the standard case $K_\mathrm{i}=0$ is shown.
%Such an integral action numerically proved to be beneficial, as it allows to obtain steady-state results comparable to those obtained with larger values of $K_\mathrm{p}$.
\section{Conclusions}
We investigated the stability and robustness properties of a continuification control strategy for a set of agents in a ring. We quantified the extent to which the approach presented in \cite{maffettone2022continuification} is affected by key realistic effects, including (i) limited sensing capabilities of the agents; (ii) presence of spatio-temporal disturbances; and (iii)  structural perturbations of their interaction kernel. In all cases, we establish the mathematical proofs of local asymptotic or bounded convergence -- the latter in the form of a residual steady-state mismatch that can be made arbitrarily small by increasing the control gain. We also reported preliminary results about the addition of a spatio-temporal integral action at the macroscopic level that can considerably reduce the steady-state error, even after the action is discretized and deployed at the microscopic agents' level. Ongoing work is aimed at analytically characterizing the effects of such an action and generalizing the approach to higher dimensions.

%We prove that, out of the nominal condition, local stability and bounded convergence can be still guaranteed, giving also an estimation of the associated basins of attraction and bounds. Several numerical simulations confirm what is found theoretically. The extension to multi-dimensional domains, {\color{blue}and  an extended theoretical analysis regarding the role of the extra integral action we briefly shown,} are subjects of ongoing work.

\bibliographystyle{IEEEtran}
% Generated by IEEEtran.bst, version: 1.14 (2015/08/26)

\end{document}